\documentclass[aps,pre,onecolumn,epsf,showpacs,graphicx]{revtex4}

\usepackage{epsfig,latexsym} 
\usepackage{amsmath,amscd}
\newcommand \eps {\epsilon}
\newcommand \lan {\langle} 
\newcommand \ran {\rangle} 
\newcommand \qt {{\tilde q}} 
\begin{document}

\title{\bf\noindent Fluctuation induced interactions between domains
in membranes}

\author{D.S. Dean and M. Manghi}
\affiliation{Laboratoire de Physique Th\'eorique, UMR CNRS 5152, IRSAMC,
Universit\'e Paul Sabatier, 118 route de Narbonne, 31062 Toulouse
Cedex 04, France}

\date{3 July 2006}

\begin{abstract}
We study a model lipid bilayer composed of a mixture of two
incompatible lipid types which have a natural tendency to segregate in
the absence of membrane fluctuations. The membrane is mechanically
characterized by a local bending rigidity $\kappa(\phi)$ which varies 
with the average local lipid
composition $\phi$. We show, in the case where $\kappa$ 
varies weakly with $\phi$, that the effective interaction between lipids
of the same type can either be everywhere attractive or can  have a
repulsive component at intermediate distances greater than the typical
lipid size. When this interaction has a repulsive component, it can
prevent macro-phase separation and lead to separation in mesophases
with a finite domain size. This effect could be relevant to certain
experimental and numerical observations of mesoscopic domains in such
systems.
\end{abstract}  
\maketitle

\section{Introduction}

At the simplest level biological membranes are modeled by 
homogeneous flexible bilayers of amphiphilic lipid 
molecules~\cite{hel,seifert}. However, in many
physical and biological situations, these membranes are inhomogeneous
on some microscopic scale. Indeed four major distinct lipid types
are typically present in mammalian cell membranes \cite{alberts}. 
It is natural to ask what may be the role of this homogeneity in the 
biological context and how it influences the mechanical 
properties of the cell. The interplay
between the lipid composition and membrane fluctuations has been
addressed in many recent studies. The local composition of the 
membrane will clearly affect its  fluctuations and local
geometry. Indeed, the coupling between membrane fluctuations and local 
composition is at the origin of the budding instability~\cite{lip,juli} 
seen in certain systems. On the other hand,  membrane fluctuations
will also influence its local composition. In this paper we will
examine how the coupling of  membrane fluctuations to local 
composition can affect the phase ordering of its component lipids. 

In previous works, the way in which the fluctuation--composition coupling is 
incorporated into the  overall free energy of system falls into 
two main classes: i) The membrane is composed of a homogeneous lipid background
with added insertions such as trans-membrane proteins and attached polymers. ii) The membrane is modeled as a multicomponent system with several
lipid types and  where the mechanical properties of the system are dependent
on the relative local concentrations of the various lipid types.

The insertions considered in models of class i) modify the membrane fluctuations via several different mechanisms. First point-like  inclusions, 
such as polymers, exert a pressure distribution on the flexible membrane. 
This involves  a coupling of the membrane composition, 
in this case the density field of the 
inclusions, to the height $h({\bf x})$ over the projected
area of the membrane. Another possible coupling is via an 
imposed  boundary condition on the height field $h$ at the boundary between
the inclusions and the membrane. For example, the contact angle at
the boundary can be taken to be fixed in order to minimize the 
hydrophobic  free energy of the insertion. This is an example
of a hard constraint. Alternatively one can introduce a
general coupling tensor, related to the orientational degrees
of freedom of the inclusions, to the local strain tensor 
$\nabla_i\nabla_j h$, from which the curvature tensor can be
extracted. This then corresponds to an energetic term which 
induces a preferred  local curvature.  In the literature several 
types of inclusions 
are considered: circular~\cite{gobrpi1,gou1,kineos}, elliptic~\cite{palu}, 
more general~\cite{dofo1,dofo2,misbah1,barfo} embedded inclusions,
as well as adsorbed cylinders~\cite{gol}. Besides introducing a tendency
for a spontaneous local curvature, which breaks the up-down symmetry
of the system, inclusions may also modify the energy associated wit
terms quadratic in the curvature tensor. For example isotropic 
inclusions may modify the local bending and Gaussian rigidities
of the membrane. In the case of two inclusions one may then explicitly
evaluate their effective interaction. To summarize, the density field of the inclusions in all these cases is coupled via: (a) $h$ in the case of insertions exerting a pressure, 
(b) an effective vectorial $\nabla_i h$ coupling in the case of
imposed boundary conditions at the inclusions frontier with the 
membrane, (c) a two tensor coupling to $\nabla_i \nabla_j h$  when there
is a locally preferred curvature tensor and finally (d) a coupling to 
$\nabla_i \nabla_j h \nabla_k\nabla_l h$, when the local bending
and Gaussian rigidities are modified by the inclusions and also when  
non-isotropic effects are present. The above 
are the most physically relevant  couplings up to quadratic order and
consequently are the most significant in systems where the 
height fluctuations are relatively small.  

In this case of models of type ii), the variation of the elastic properties of the membrane is more continuous than in the case of inclusions. If one
neglects the possibility of non-isotropic effects, the most natural
parameters which will vary with local lipid composition are the
bending rigidity $\kappa$, the Gaussian rigidity $\overline{\kappa}$
and the spontaneous local curvature $c$. For instance a  concentration 
dependent spontaneous curvature is considered 
in~\cite{leibler,tani,saxena,misbah2}. Linear perturbations 
to both the bending rigidity  and the spontaneous curvature are 
studied in~\cite{netz,nepi}. In~\cite{gobrpi1}, linear perturbations to 
the bending and Gaussian rigidity were considered, the 
interaction arising in this case is proportional to $1/r^4$ and the 
prefactor is given by the product of  coefficients of  the 
linear deviations from the average value of $\kappa$ and $\overline{\kappa}$. 
The induced interaction may thus be attractive or repulsive depending on the
sign of these coefficients. Of course models of type (i), with discrete 
inclusions, can be described by models of type (ii) when it makes sense to
take a continuum limit for the inclusions. This limit will be valid for
inclusion sizes which are comparable to the microscopic length scale
of the membrane, that is to say the lateral lipid size.  

All of the studies mentioned above  assume zero surface tension. 
However, the study of membranes under tension sheds light upon 
the physics of biological membranes which are not truly at 
equilibrium but under external constraints or perturbations.
The surface tension can be due to electrostatic interactions with 
the aqueous solvent or due to the presence of molecular protrusions. 
Furthermore, the external action of laser tweezers on a vesicle 
attracts  phospholipids an puts the membrane
under tension. This leads to interesting phenomena such as  pearling
instabilities~\cite{barziv}. It has been theoretically shown that the
presence of surface tension can induced a \emph{repulsive} interaction
between inclusions of the same type~\cite{sens,sens2}. The model used
in~\cite{sens,sens2} is of type i) and is based on a linear 
coupling of the inclusion to the height of the membrane, for
example to model  the local pressure exerted by an attached polymer. The interaction is sensitive to the strength of membrane-inclusion coupling. In this
system, the up-down symmetry of the membrane is clearly broken by the
linear coupling. Indeed, in many biological situations the up-down symmetry of the
membrane is clearly broken for instance by different compositions in the top and bottom leaves or by the presence of conical trans-membrane inclusions. However, it is interesting to ask if the presence of surface tension can also lead to repulsive interaction between domains, with similar lipid composition, even when the up-down symmetry is conserved. 

The physics of phase separation may play an important role in biological 
systems. It has been experimentally shown that erythrocyte membranes which contain many different lipid types form immiscible two dimensional liquids, which are very close to the miscibility critical point~\cite{keller}. The resulting thermodynamic
forces can affect the mechanical properties of the membrane
and in particular its shape. However, in turn the fluctuations 
will also affect the distribution of the components in the
membrane. As an example, a long-range fluctuation mediated
repulsion between inclusions, combined with a short-range van der Waals
attraction, could lead to the formation of mesoscopic
domains~\cite{tani,sear} of the inclusions. It has also been shown that
the presence of a surface tension modifies the effective interactions between
conical inclusions \cite{wkh}, inclusions of the same type are always 
repelled but oppositely orientated inclusions interact attractively at
long distances and then repel at shorter distances. This is in contrast
to the case where there is no tension when all interactions are always
repulsive.

In this paper, we consider a two-component bilayer with the up-down
symmetry and, in general, with a non-zero surface tension. We show that for
certain variations of the bending rigidity and the local surface energy
(the composition independent component of which can be interpreted as
a surface tension) with the local composition in lipids, a fluctuation induced
lipid-lipid repulsive interaction can appear between domains
of similar composition. This, together with a
short-range van der Waals attraction, can induce the formation of
mesophases. In the scheme of previous models, our model falls into the class
of  type ii) above  and our cumulant expansion method is similar to  that
used in \cite{gobrpi1} and \cite{netz,nepi}. In our study we add 
an non-zero surface energy, as  in ~\cite{sens,sens2}, but 
where this local surface energy  fluctuates with the local lipid composition.

The paper is organized as follows. 
In Section~II we present our field theoretical model. In section~III
using a cumulant expansion for small height fluctuations we
calculate the induced interaction, this rather technical section
may be skipped by a reader interested only in the physical
consequences of the calculation.  In section~IV the general physical
properties and asymptotic behavior of this effective interaction are
discussed. Section~V is devoted to a description of the results
which are compared to previous studies. In addition we suggest a
possible experiment where the effects predicted here could possibly be seen.

\section{Field-theoretical formulation}

We consider a model membrane with two lipid types $A$ and $B$ and
where the top and bottom leaves have the same lipid composition. In
the most frequent case, at least whenever van der Waals interactions
are dominant, it is energetically favorable for lipids of the same
type to be adjacent. In this case, we can write down a typical
attractive energy per site $E=\chi \phi_A\phi_B$ where $\phi_A$ and
$\phi_B$ are the liquid volume fraction of lipid $A$ and $B$ and
$\chi>0$ is a Flory parameter related to electronic polarizabilities
of both molecules. We  will consider a coarse-grained model  
for a field $\phi$ related to the local surface fraction of the two lipid
types, i.e. $\phi=\phi_A-\phi_B$, which in the absence of surface 
fluctuations  exhibits a continuous phase transition
at sufficiently low temperatures. The theory is then described by the
Ginzburg-Landau Hamiltonian~\cite{tani}
\begin{equation}
H^I[\phi] = \int \sqrt{g}d^2{\bf x}\ \left[{J\over 2}
g^{ij}\nabla_i\phi\nabla_j \phi + V(\phi)\right],\label{eqm0}
\end{equation}
which is written in a covariant form which ensures the independence of the 
energy from the choice of the two dimensional coordinate system
denoted by ${\bf x}$. 
The parameter $J$ is positive and related to the Flory
parameter $\chi$, it is a ferromagnetic interaction and
energetically favors lipids of the same type being next
to each other. The potential $V(\phi)$ fixes the two characteristic
values of $\phi$ and the global composition via chemical potential like
terms.  As the potential $V$ appears in $H$ simply integrated over the 
area of the membrane, it  can be interpreted as a
\textit{composition dependent contribution to the surface energy} of the 
membrane. Indeed, the constant part of $V$ which is $V(0)$ can be 
interpreted as
a surface tension because it is coupled to the total physical
area of the membrane $\int \sqrt{g}d^2{\bf x}$. 
The term $V(0)$ can thus be used as a Lagrange multiplier to fix the 
physical membrane area. As mentioned above, $V$ will have a $\phi$ dependence
as in the usual Landau models for phase separating systems. 
As in standard Landau theory, we will assume that $V$ is a single well at 
high temperature and a double well at low temperature. This means that 
the system on a plane will exhibit  a continuous phase transition.
At the mean field level, this transition occurs when when the mass associated to this field theory, given by $M^2_0= V''(\phi_0)$, vanishes (where $\phi_0$ is the homogeneous 
mean field solution). This transition exhibits a divergent 
correlation length and corresponds to a macro-phase separation which occurs at a critical temperature $T=T_c$. 

In the above, the metric of the membrane surface is denoted by $g_{ij}$ and in the Monge gauge it is given by
\begin{equation}
g_{ij} = \delta_{ij} + \nabla_i h \nabla_j h
\end{equation}
where $h$ is the height of the surface above the projection 
plane whose area we will denote by $A$. 
The term $g$ denotes the determinant of $g_{ij}$ and is given by
\begin{equation}
g = 1 + (\nabla h)^2.
\end{equation} 
Hence the Hamiltonian given by Eq.~(\ref{eqm0}) already implicitly
includes a coupling between the local composition,
as encoded by $\phi$, and the membrane fluctuations, as
encoded by $h$. The interface energy or line tension, which corresponds to the
term quadratic in the gradient, is written as  to ensure
covariance; $g^{ij}$ is the inverse of $g_{ij}$ and is given by
\begin{equation}
g^{ij} = \delta^{ij} - {\nabla_i h \nabla_j h\over 1 + (\nabla h)^2}.
\end{equation} 
Here we note that the fact that one should use the covariant form of the line tension is often forgotten in the literature. To lowest, i.e. quadratic order in the fluctuations $h$, one has
\begin{equation}
H^I[\phi,h] = \int d^2{\bf x}\ \left[{J\over 2} (\nabla\phi)^2 +
V(\phi)\right] + \int d^2{\bf x}\ \left[{J\over 4} (\nabla\phi)^2
(\nabla h)^2 - {J\over 2} (\nabla\phi \cdot \nabla h)^2 + {1\over
2}(\nabla h)^2V(\phi)\right]
\label{HIc}
\end{equation}

We now take into account the elastic energy of the membrane so the total Hamiltonian of the system is  given by
\begin{equation}
{\cal H}[\phi,h] = H^I[\phi,h] + H^S[\phi,h].\label{H}
\end{equation}
The Hamiltonian for surface fluctuations will be taken to be
\begin{equation}
H^S[\phi,h] = {1\over 2}
\int d^2{\bf x}\ \kappa(\phi) \left(\nabla^2 h\right)^2  \label{eq2}
\end{equation}
which is the  simplest Helfrich Hamiltonian for surface 
fluctuations~\cite{hel} and correspond, strictly speaking, to the first term in an $\nabla^2 h$-expansion of the mean curvature~\cite{seifert}. This Hamiltonian corresponds to  a bending energy with local bending rigidity which depends on the local composition characterized by $\phi$. The two dimensional membrane system is assumed to have no
spontaneous curvature and thus has an up-down symmetry. More generally, 
one could also include a composition dependence on  the Gaussian rigidity, 
the contribution coming from this term would then  cease to be a topological
invariant and should strictly be included.

The effective partition function in the
presence of membrane fluctuations is given by
\begin{equation}
Z = \int d[\phi]d[h] \exp(-\beta {\cal H}) \label{Z}
\end{equation}
where $\beta^{-1}=k_BT$ is the thermal energy scale. 
We recall that $A$ is the projected area of the membrane,
the physical area of the membrane is denoted by $A+\Delta A$, where 
$\Delta A$ is often called the excess area. For typical biological membranes,  
$\Delta A/A$  is small, of the order of a few percent, and we will thus 
legitimately assume, in the rest of the paper, that height fluctuations 
are small compared  to the typical length scale of the system. 

\section{Calculation of the fluctuation induced interaction}
In this section we  explicitly calculate the fluctuation induced 
interaction to second order in the cumulant expansion.

In the high temperature regime, lipids form a mixed phase characterized
by a homogeneous and uniform composition $\phi_0$, with fluctuations
$\psi$ about $\phi_0$. In an ensemble where the average value
of $\phi$ is fixed write $\phi=\phi_0+\psi$ where
$\phi_0=A^{-1}\int d^2{\bf x}\,\phi({\bf x})$. Consequently in this case, 
we have $\int d^2{\bf x}\,\psi({\bf x})=0$.    
By assuming that $\kappa$ and
$V$ behave continuously around $\phi_0$, we expand the total
Hamiltonian~(\ref{H})  up to
$\mathcal{O}(\psi^2)$ in the fluctuations. This leads to
\begin{eqnarray}
{\cal H}[\phi_0,\psi,h] &=& AV(\phi_0) +H^I_0[\phi_0,\psi] +
H^S_0[\phi_0,h] + \Delta H[\phi_0,\psi,h],\\ H^I_0[\psi] &=& \frac12
\int d^2{\bf x}\ \left[J (\nabla\psi)^2 +  
V''(\phi_0) \psi^2 \right],\\
H^S_0[h] &=&\frac12 \int d^2{\bf x}\ \left[\kappa(\phi_0)
\left(\nabla^2 h\right)^2 + V(\phi_0) \left(\nabla
h\right)^2\right].
\label{HS0}
\end{eqnarray}
When the term proportional to 
$V(\phi_0)$  is included in the surface Hamiltonian $H_0^S$, as we have
chosen to do above, $V(\phi_0)$ can be interpreted as an effective 
elastic energy. However, because
it is constant, $V(\phi_0)$ can be interpreted as an effective  surface 
tension. The part of the Hamiltonian which we will treat perturbatively is
\begin{eqnarray}
\Delta H[\phi_0,\psi,h] &=& \frac12\int d^2{\bf x}\
\left\{J\left[{1\over 2} (\nabla\psi)^2 (\nabla h)^2 - (\nabla\psi
\cdot \nabla h)^2\right] + \left[V'(\phi_0)(\nabla
h)^2+\kappa'(\phi_0)(\nabla^2 h)^2\right]\psi\right. \nonumber \\ &+&
\left.\frac12\left[V''(\phi_0)(\nabla
h)^2+\kappa''(\phi_0)(\nabla^2 h)^2\right]\psi^2\right\}.
\end{eqnarray}
The scheme of the calculation is just slightly different in the case where the
value of $\phi_0$ is allowed to fluctuate but nothing intrinsically changes. 

We perform a
cumulant expansion in the partition function~(\ref{Z}) as follows
\begin{eqnarray}
Z &\approx&  Z^S_0\exp(-\beta A V(\phi_0)) \int
d[\psi]\exp(-\beta H^I_0) \left[1-\beta \lan \Delta H\ran^S +
\frac{\beta^2}2 \lan(\Delta H)^2\ran^S\right]\\ &\approx& \int
Z^S_0 \exp(-\beta A V(\phi_0)) \int
d[\psi]\exp\left(-\beta {\cal H}_{\mathrm{eff}}[\phi_0,\psi]\right)
\label{eqpp}
\end{eqnarray}
where $Z^S_0=\int d[h]\exp(-\beta H^S_0)$ and $
\lan\mathcal{O}\ran^S=(Z^S_0)^{-1}\int d[h]\mathcal{O}\exp(-\beta
H^S_0)$. The cumulant expansion at this order is clearly exact to
 $\mathcal{O}(\psi^2)$. 
The effective interaction at this order is thus given by
\begin{equation}
{\cal H}_{\mathrm{eff}}[\phi_0,\psi]=H_0^I[\phi_0,\psi] + \lan \Delta
H\ran^S_c - \frac{\beta}2 \lan(\Delta H)^2\ran^S_c
\end{equation}
where the subscript $c$ indicates that it is the connected part of the
correlation function.  

Note that only the first term in the cumulant
expansion can lead to a quadratic terms in $\nabla\psi$, however this 
term can be seen to be zero by the following:
\begin{eqnarray}
\lan \frac12(\nabla\psi)^2 (\nabla h)^2 - (\nabla\psi \cdot \nabla
h)^2\ran^S_c &=& \frac12(\nabla\psi)^2\lan (\nabla h)^2\ran^S_c -
\nabla_i \psi\nabla_j\psi \lan \nabla_i h\nabla_j h \ran^S_c \nonumber
\\ &=& \frac12 (\nabla\psi)^2\lan (\nabla h)^2\ran^S_c - \nabla_i
\psi\nabla_j\psi \ \delta_{ij}\  \frac12 \lan (\nabla h)^2\ran^S_c = 0,
\end{eqnarray} 
where we have appealed to the isotropy of the system. There is therefore
no renormalization of the coupling $J$. Also in the 
first term of the cumulant expansion, terms linear in $\psi$ cancel 
by definition of $\psi$ (as they are integrated against a constant
by isotropy) and the remaining terms yield 
$\lan \Delta H\ran^S_c = \frac12\int d^2{\bf x} M_1^2\psi^2$
where the mass $M_1$ is given by
\begin{equation}
M_1^{2} = \frac12\left[V''(\phi_0) \lan (\nabla h)^2\ran^S_c +
\kappa''(\phi_0)\lan (\nabla^2 h)^2\ran^S_c \right].
\end{equation}
Again to quadratic order in $\psi$, the  second order term in  
the cumulant expansion yields
\begin{equation}
-\frac{\beta}2 \lan(\Delta H)^2\ran^S_c= \frac12 \int d^2{\bf x}d^2{\bf
 y} \psi({\bf x})U({\bf x} -{\bf y})\psi({\bf y})
\end{equation}
where
\begin{equation}
U({\bf x} -{\bf y}) = -{\beta\over 4} \left[ {\kappa'}^2(\phi_0) \lan
\left(\nabla^2 h({\bf x})\right)^2 \left(\nabla^2 h({\bf
y})\right)^2\ran^S_c + 2 V'(\phi_0)\kappa'(\phi_0) \lan
\left(\nabla^2 h({\bf x})\right)^2 \left(\nabla h({\bf
y})\right)^2\ran^S_c + {V'}^2(\phi_0) \lan \left(\nabla h({\bf
x})\right)^2\left(\nabla h({\bf y})\right)^2\ran^S_c\right].
\end{equation}
The potential $U({\bf x} -{\bf y})$ is non-local and characterizes the
induced interaction mediated by height fluctuations. The various
connected correlation functions above are evaluated as
\begin{eqnarray}
\left\langle (\nabla^2 h({\bf x}))^2 (\nabla^2 h({\bf y}))^2
\right\rangle_{c}^S &=& {2\over \beta^2} (\nabla^4 G({\bf x}-{\bf
y}))^2 \nonumber \\ \left\langle (\nabla h({\bf x}))^2 (\nabla h({\bf
y}))^2 \right\rangle_{c}^S &=& {2\over \beta^2} \sum_{ij}
(\nabla_i\nabla_j G({\bf x}-{\bf y}))^2 \nonumber \\ \left\langle
(\nabla^2 h({\bf x}))^2(\nabla h({\bf y}))^2 \right\rangle_{c}^S &=&
{2\over \beta^2} (\nabla \nabla^2 G({\bf x}-{\bf y}))^2.
\end{eqnarray}
Here the Green function $G$ is given by
\begin{equation}
G = (\kappa(\phi_0) \nabla^4 -V(\phi_0) \nabla^2)^{-1} =
{1\over V(\phi_0)}( G_0-G_m),
\end{equation}
where
\begin{equation}
\xi=\frac1{m} = \sqrt{\frac{\kappa(\phi_0)}{V(\phi_0)}}.
\end{equation} 
The intrinsic length $\xi$ is usually in the range 10--100~nm for
biological membranes. The Green function $G_0$ is $G_0=-\frac1{2\pi}\ln\left|\frac{\bf x}{L}\right|$ ($L$ is an arbitrary length) and
$G_m$ is the Yukawa interaction given by
\begin{equation}
-\nabla^2 G_({\bf x}) + m^2 G_m({\bf x}) = \delta({\bf x}).
\end{equation}
In two dimensions one has
\begin{equation}
G_m({\bf x}) = \frac1{2\pi }K_0(m|{\bf x}|),
\end{equation}
where $K_0(x)$ is the Bessel function of the second kind of order 0.
Using these results we find that
\begin{equation}
U({\bf x}) = B \delta({\bf x}) + v(|{\bf x}|).
\end{equation}
The first term of the r.h.s. is short-ranged with
$B=-(m^4/2\beta)[\kappa'(\phi_0)/
\kappa(\phi_0)]^2[\delta(0)-2m^2G_m(0)]$ and needs to be regularized via 
an ultra-violet cut-off $\Lambda=2\pi/a$ corresponding to a microscopic
length scale $a$ which would be of the order of the distance between lipid heads. 
The second term $v$ of the r.h.s. above is
the \textit{long-range} induced interaction and is independent of 
the ultra-violet cut-off. It is
given by
\begin{eqnarray}
v(r) &=&
\frac{m^4}{2\beta}\left[\frac{\kappa'(\phi_0)}{\kappa(\phi_0)}\right]^2
\nu_{\alpha}(u),
\label{v} \\ \nu_{\alpha}(u) &=& - \frac1{4\pi^2} \left\{
K_0^2(u) + 2\alpha K^{\prime 2}_0(u) + \alpha^2 \left[ \left(K^{\prime
\prime}_0(u) -{1\over u^2}\right)^2 +{1\over
u^2}\left(K^{\prime}_0(u)+{1\over u}\right)^2\right]\right\} 
\end{eqnarray}
where $u = m|{\bf x}|=mr$ and
\begin{equation}
\alpha =
\frac{V'(\phi_0)}{V(\phi_0)}\frac{\kappa(\phi_0)}{\kappa'(\phi_0)}.\label{defa}
\end{equation}

\section{Properties of the fluctuation induced interaction}

Here we discuss the  features of the interaction 
derived in section~III. 
First, the strength of this interaction is polynomial in $m$ and 
therefore reducing the local surface or elastic energy and increasing 
the local bending rigidity reduces
considerably the interaction energy. This same reduction however  
also increases the range of the interaction. Secondly, the interaction 
strength is set by $k_B T$, which means that this interaction has an 
entropic origin.

If $\alpha$ is positive, we see that the interaction between areas
having the same lipid type (with the same sign of $\psi$) is always
attractive. To see this, we note that in the expression for $v(u)$,
Eq.~(\ref{v}), the coefficients of $\alpha^0$, $\alpha$ and $\alpha^2$ 
are functions of $u$ which are positive and monotonically decreasing
hence yielding an attractive interaction at all distances. 
In a ferromagnetic analogy where the lipid types are characterized by a field 
$\phi$ having the values concentrated about $\pm 1$, this corresponds 
to a long-range ferromagnetic interaction which enhances the short-range 
one already present. However when $\alpha$ is negative, $v$ now has a 
repulsive component, corresponding to the coefficient of $\alpha$, 
which could prevent macro-phase separation. 

For large $u$ we have~\cite{abram}
\begin{equation}
K_0(u) \approx \sqrt{\frac{\pi}{2 u}}\exp(-u)
\end{equation}
and thus at large distances $v$ behaves as
\begin{equation}
v(r) \approx
-\frac{k_BT}{4\pi^2}\left[\frac{V'(\phi_0)}{V(\phi_0)}\right]^2\frac1{r^4}.
\label{approxlarger}
\end{equation} 
This $1/r^4$ interaction, which is always attractive, is typically 
seen between inclusions of the same type  in membranes without 
surface tension and is found in many
of the studies discussed in the Introduction. 
However, this $1/r^4$ attraction  does not  have the the  same physical 
origin as in previous studies because, as can be seen by examining the 
prefactor, it is generated solely by the fluctuations of the surface energy.  

For small $r$ we find
\begin{equation}
v(r) \approx -\alpha \frac{k_BT}{8 \pi^2}
\left[\frac{\kappa'(\phi_0)}{\kappa(\phi_0)}\right]^2 \frac{m^2}{r^2}
+\mathcal{O}(\ln^2r) \approx - \frac{k_BT}{4 \pi^2}
\frac{\kappa'(\phi_0)V'(\phi_0)}{\kappa^2(\phi_0)} \frac1{r^2}
+\mathcal{O}(\ln^2r)
\end{equation}
which is again attractive if $\alpha >0$ but is repulsive when $\alpha
<0$ (or $\kappa'(\phi_0)V'(\phi_0)<0$). In this last case, the
overall interaction is somewhat frustrated: it is attractive at very 
short length scales (of the order of the microscopic length scale) due to van 
der Waals interactions (in our model represented by the local ferromagnetic interaction), together with a longer range membrane mediated repulsion
over intermediate length scales, before becoming 
attractive at longer length scales. One can can suppose that the
occurrence of these attractive and repulsive interactions can prevent
macroscopic phase separation and lead to mesoscopic domains.

\section{Microphase separation}

The final effective quadratic Hamiltonian can now be written as 
\begin{equation}
{\cal H}_{\mathrm{eff}}[\phi_0,\psi] = \frac12 \int d^2{\bf x}\
\left[J(\nabla\psi)^2 + M_2^2 \psi^2 \right] + \frac12 \int d^2{\bf
x}d^2{\bf y} \psi({\bf x}) v({\bf x} -{\bf y})\psi({\bf y}),
\end{equation}
where the corrected mass is given by $M_2^2= M_0^2 +M_1^2 +B$. It is
important to note that this mass depends on the microscopic cutoff $a$
since $M_1^2$ includes the expectation values $\lan (\nabla^2 h)^2\ran^S$
and $\lan (\nabla h)^2\ran^S$ which diverge and must thus
be regularized and a similar regularization is needed to evaluate
$B$. As already explained in the
Introduction, we consider systems such that, 
in the absence of height fluctuations, 
when $M_0\to 0$ the system exhibits a second order phase transition with 
diverging correlation length $\ell_0 = \sqrt{J} /M_0$. 
In Fourier space, we find
\begin{equation}
{\cal H}_{\mathrm{eff}}[\phi_0,\psi] = {1\over 2 (2\pi)^2} \int d^2{\bf q} \left[J
q^2 + M_e^2 +w(q) \right] {\tilde \psi}({\bf q}){\tilde \psi}({\bf
-q}),
\end{equation}
where $q = |{\bf q}|$ and the Fourier transform and its inverse are
defined by
\begin{eqnarray}
f({\bf x}) &=& \int \frac{d^2{\bf q}}{(2\pi)^2} {\tilde f}({\bf
q})\exp(i{\bf q}\cdot {\bf x}),\\ {\tilde f}({\bf q}) &=& \int d^2{\bf
x} f({\bf x})\exp(-i{\bf q}\cdot {\bf x}).
\end{eqnarray}
In the Fourier representation the non-local part of the interaction
is given by $w(q)= {\tilde v}(q)-{\tilde v}(0)$ and $M_e^2=M^2_2+ {\tilde v}(0)$ thus gives  effective mass for the theory.

The stability of the homogeneous solution against phase separation is
determined by the lipid-lipid correlation function in Fourier
space $\lan {\tilde \psi}({\bf q}){\tilde \psi}({\bf q}')\ran =(2\pi)^2 \delta({\bf q}+{\bf q}')S({\bf q})$ where the structure factor is
\begin{equation}
S({\bf q}) = \frac{k_BT}{J q^2 + M^2_e + w(q)}\label{S}
\end{equation}

\begin{figure}[h]
\includegraphics[height=6cm]{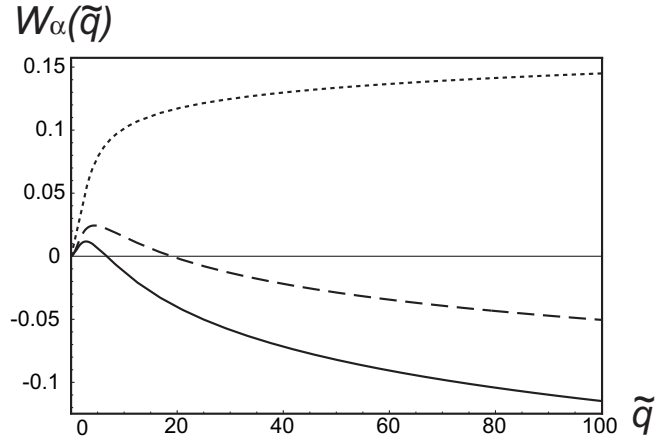}
\caption{Plot of $W_{\alpha}(\qt)$ for various values of $\alpha$:
$0.05$ (dotted line), $-0.1$ (broken line) and $-0.15$ (solid line).}
\label{W}
\end{figure}

Defining 
\begin{equation}
\eps=\left[\frac{\kappa(\phi_0)}{\kappa'(\phi_0)}\right]^2,
\end{equation}
we find three dimensionless parameters in the structure factor
\begin{equation}
{\tilde S}(\qt)=\frac{S(\qt)}{2\eps\xi^2}=\frac1{2\eps\beta J
\left[(\xi /\ell_e)^2+\qt^2\right] + W_{\alpha}(\qt)} \label{sf}
\end{equation}
where $\qt=\xi q$, which are $\alpha$, $\eps\beta J$ and $\ell_e^2=J/M^2_e$. The Fourier transform of the dimensionless potential is
\begin{equation}
W_{\alpha}(\qt) = 2\pi\int_0^\infty u du \left[J_0(\qt u) -1\right] \nu_{\alpha}(u) ,
\end{equation}
where $J_0(x)$ is the Bessel function of the first kind of order 0. A
divergence of the structure factor, Eq.~(\ref{sf}) at $\qt=0$, while 
$S$ remains finite for $\qt\neq 0$ signals a macro-phase separation. 
In this case, since  we find $S(0)=\ell_e^2/(\beta J)$, the phase separation 
occurs  when the correlation length $\ell_e \to \infty$ and this 
corresponds to the case where the induced interaction does not change 
the nature of the transition but only changes where it occurs. We show the
behavior of $W_{\alpha}(\qt)$ and ${\tilde S}(\qt)$ for different
values of $\alpha$ in Fig~\ref{W}. When $\alpha \ge 0$,
$W_{\alpha}(q)$ and its first derivative are always positive, and the
only maximum of the structure factor occurs at $\qt=0$.

However, when $\alpha <0$ we see that for particular values of
$\eps\beta J$ and $\xi/\ell_e$, ${\tilde S}(\qt)$ reaches a maximum at
an \emph{intermediate} values of $\qt$ and this maximum can even
diverge at a non-zero wave-vector $\qt^*$. In this case, the
homogeneous solution becomes unstable before $\ell_e \to \infty$ which
leads to the formation of \textit{mesophases} (mesoscopic phase
separation) with a finite characteristic length scale given by
$\xi/\qt^*$. Note that at large values of $\qt$, we have
$W_{\alpha}(\qt)\sim \alpha \ln \qt$ but this  short-range component of the 
induced interaction is dominated by the short-range van der 
Waals interaction term whose strength is controlled by $J$, and $S(\qt)$ 
ultimately decreases as $1/\qt^2$ for large $\qt$.
The maximum of the structure factor diverges for an intermediate
wave-vector $\qt^*$ which is implicitly defined by the two following
equalities
\begin{eqnarray}
W'_{\alpha}(\qt^*) + 4\eps\beta J\qt^* &=& 0 \label{minimum}\\
W_{\alpha}(\qt^*) + 2\eps\beta J\left[\qt^{*2} + (\xi/\ell_e)^2\right]\label{zero}
&=& 0
\end{eqnarray}
i.e. when the parabola $-2\eps\beta J[\qt^2 + (\xi/\ell_e)^2]$ is
tangent to $W_{\alpha}(\qt)$. Given the number of parameters in our theory
the evaluation of a complete phase diagram is unfeasible, however the \
fundamental question we wish to address is whether there is a macro-phase
or micro-phase separation. To do this we can examine the structure factor
at the point where the ${\bf q}=0$ mode becomes unstable, that is to say
where the effective mass $M_e =0$. This is thus equivalent to examining 
temperatures $T_c$ which are critical in the true sense. If the modes
${\bf q} \neq 0$ are stable at $T_c$ then we expect to see the macro-phase 
separation. However if at $T_c$ there is already a mode $q\neq 0$ 
which is unstable then  a micro-phase separation must have already 
occurred at a temperature $T> T_c$. Thus, without having to specify the full 
theory, we can identify when  a macro-phase separation is converted to a 
micro-phase one due to  coupling between membrane fluctuations and its 
composition.

\begin{figure}[h]
\includegraphics[height=8cm]{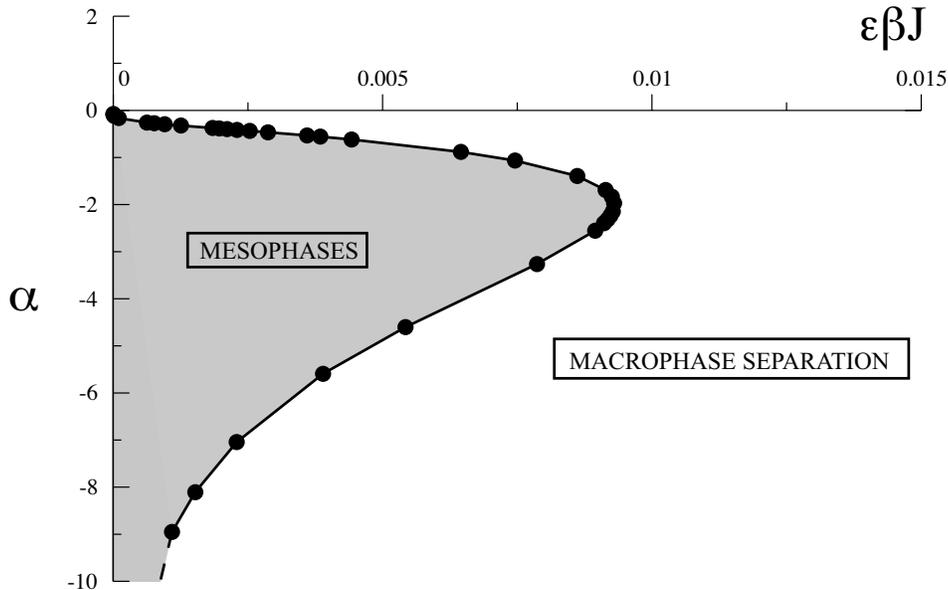}
\caption{Theoretical phase diagram of the bi-lipidic fluctuating membrane in the plane ($\epsilon\beta J$, $\alpha$) at the onset of the macro-phase separation, i.e. at the critical temperature $T_c$. The gray region corresponds to the parameter range for which a phase separation occurs at a non-zero wave vector $\qt^*$ leading to the formation of mesophases instead of a macrophase separation.}
\label{diagram}
\end{figure}

Since we are interested in the behavior
of the structure factor when approaching the macro-phase transition
($\ell_e \to \infty$); we calculate the onset of the micro-phase
separation given by Eqs.~(\ref{minimum})--(\ref{zero}) for
$\xi/\ell_e=0$ i.e. at the critical temperature $T_c$. The result is
shown in Fig.~\ref{diagram}: the gray region corresponds to the region
of the phase diagram ($\epsilon\beta J$, $\alpha$) where mesophases
appears whereas the white region corresponds to macro phase
separation. The solid line corresponds to the solution of
Eqs.~(\ref{minimum})--(\ref{zero}) (the dots are the exact
solutions). This "phase diagram" is plotted at a fixed lipid
composition $\phi_0$ and fixed temperature $T=T_c$ corresponding to
the critical point in the ($\phi$, $T$) space. It is important to note
that when the temperature deviates from the critical
temperature, the correlation length $\ell_e$ becomes finite (but very
large), and the gray region delimited by the solid line
shrinks. However, in the case where, for given lipid types, the
parameters $\epsilon\beta J$ and $\alpha$  lie in the gray region,
mesophases appear before the macro-phase separation at a temperature
$T>T_c$. In the extremal situation where we are far from the
macro-phase separation the region of the phase diagram corresponding to
mesophases disappears completely.

For parameter values belonging to the gray region of the diagram, the
structure factor ${\tilde S}(\qt)$ diverges before $\qt=0$ at a finite
value $\qt^*$. When moving along the solid line starting at the
origin, the value of $\qt^*$ decreases until we reach the point
($\epsilon\beta J=3.8\,10^{-3}$, $\alpha=-0.557$) where $\qt^*=2.095$
which is the smallest value and corresponds to a characteristic length
scale of 5--50~nm for the mesophases. Then $\qt^*$ increases again when $|\alpha|$
increases.

On the experimental side, the parameters $\alpha$ and $\epsilon\beta J$ are not easy to determine. The phenomenological parameter $J$ coming from the Landau-Ginzburg theory, is somehow related to the van der Waals attractions between lipids and varies roughly as $1/T$ such that $\beta J$ is fixed for given lipid types and independent on temperature. Although we do not know exactly the value of this parameter, we can assume $\beta J\sim1$. For lipidic vesicles made of a mixing of two very different lipids such as DMPC/cholesterol (a long lipid and a short one), the curvature modulus has been experimentally measured $\kappa_0\simeq 50\, k_BT$ ($26\, k_BT$ with DMPC alone) and increases when the proportion of cholesterol increases up to $250 \,k_BT$ with 50\% of cholesterol~\cite{meleard,duwe}. Hence we find $\epsilon\beta J\sim 0.01$ which means that the mesophases region of the phase diagram can be experimentally reached in such systems.   

Finally, in the mesophase region, the denominator of $S(q)$ given by Eq.~(\ref{S}) becomes negative and the calculation of the preceding sections, based on quadratic fluctuations of $\psi$ is no longer valid (the modes $q \approx q^*$ are unstable). The nature of the resulting stable mesophase requires further analysis to determine it and this is beyond the scope of the  present paper.

\section{Discussion and conclusions}

In this paper we have shown that the coupling of membrane composition
via a composition dependence on the local surface energy and bending
rigidity can alter the phase diagram of a membrane composed of a mixture
of different lipids. Indeed, depending on the physico-chemical properties of the lipids (for instance by modifying the length of their hydrophobic tail), the membrane can exhibits a micro-phase separation leading to the formation of so-called mesophases at a temperature $T>T_c$, i.e. before an eventual macro-phase separation. In the previous works where surface tension was considered
the composition--fluctuation coupling was linear and the up-down symmetry
of the system thus broken. 

The long-range part of the interaction given by Eq.~(\ref{v}) behaves as $-1/r^4$ for large  $r$. This interaction has the same behavior as that found
between inclusions in several models where surface tension is
not present. For instance in tensionless membranes one finds the effective
pairwise interaction~\cite{gou1,palu}
\begin{equation}
V_C(r) = -k_BT\, 6\pi^2\left(\frac{r_0}{r}\right)^4\label{casimir}
\end{equation}
between circular inclusions with the up-down symmetry, 
where $r_0$ is the radius of the inclusions. 
The long-range part of the interaction found in Eq.~(\ref{v})--(\ref{approxlarger}) is also proportional to the thermal energy but it is solely due to fluctuations  in the surface or elastic energy. 

A few works focused on the effect of the surface tension on 
fluctuation induced
interactions~\cite{sens,sens2,sepangi,misbah2}. Calculating the
potential between two circular inclusions which locally apply a
pressure on the membrane, Evans \textit{et al.} found an interaction
which is everywhere repulsive~\cite{sens} between inclusions of the same
type and is given by
\begin{equation}
\Phi(r)=\frac{\zeta_1\zeta_2}{2\pi\gamma}K_0(mr),\label{Phi}
\end{equation}
where $\zeta_i$ is related to the force distribution of inclusions $i$
acting on the membrane surface and $\gamma$ is the surface tension. 
Here again, this interaction is
different from Eq.~(\ref{v}) in origin but has some similar features: it is present for membranes under tension and is repulsive with a
typical range of $\xi\simeq 30$~nm for biological membranes. Our model
is very different, it does not assume any pressure
distribution acting on the membrane but relies on the behavior of
$\kappa(\phi)$ and $V(\phi)$ close to a liquid-liquid
immiscibility critical point. This proximity to a liquid-liquid 
immiscibility critical point in a real biological context is 
supported by beautiful experiments on mono-layers made of lipids 
extracted from erythrocytes~\cite{keller}.

In this study we have seen that the induced interaction only has a repulsive 
when component when $\alpha<0$. Qualitatively, this means that the 
signs of $\kappa'(\phi_0)$ and $V'(\phi_0)$ are opposite: 
when a region is locally enriched for instance in lipid $A$, bending rigidity 
increased ($\kappa'(\phi_0)>0$) whereas the effective surface tension 
decreases ($V'(\phi_0)<0$).  Let us consider for a moment 
the mean-field theory 
where one neglects  the fluctuations $\psi$ about $\phi_0$.
Consider an incompressible membrane  which is constrained to have a constant projected area, for example a membrane supported by a frame.
Also let the membrane exchange lipid species with the bulk solution
around it \cite{fapi}. 
The mean-field free energy as a function of $\phi_0$ is given 
from Eq. (\ref{eqpp}) as
\begin{equation}  
{F(\phi_0)\over A} = V(\phi_0) +{1\over 4\pi\beta}
\int_0^{\Lambda} kdk \ln\left[\kappa(\phi_0)k^4 + V(\phi_0) k^2\right]
\end{equation}
This mean-field free energy must be  regularized by the ultra-violet
cut-off $\Lambda$. As the membrane is in  a solution containing a 
reservoir of lipid species, $\phi_0$ is not fixed but is thermodynamically selected so as to minimize the mean-field free energy. In this case, in our previous treatment we should have thus included a term $V'(\phi_0)\psi$ in the expression  for $H_0^{I}$, however this term can be seen to cancel exactly with the first term of the cumulant 
expansion, which in this case is also now no longer zero. 
The part of the
free energy $F^*$ which varies with $\phi_0$ is given by
\begin{equation}
{F^*(\phi_0)\over A} = V(\phi_0) +{1\over 8\pi\beta}
 \left[\Lambda^2 \ln\left(\kappa(\phi_0) + {V(\phi_0)\over \Lambda^2}\right)
+ {V(\phi_0)\over \kappa(\phi_0)} \ln\left({\kappa(\phi_0)\Lambda^2\over V(\phi_0)} +1 \right)\right].\label{feme}
\end{equation}
The calculation carried out in this paper is valid for small surface 
fluctuations, a way of ensuring that the fluctuations are small is
by choosing a very stiff membrane. This can be ensured by taking 
$\kappa(\phi)$ large. The equation minimizing $F^*(\phi_0)$ can be
expressed as
\begin{equation}
V'(\phi_0)\left[ 1 + {1\over 4\pi \beta}\int_0^\Lambda kdk\ {1\over 
\kappa(\phi_0) k^2 + V(\phi_0)}\right]  + {\kappa'(\phi_0) \over 4\pi \beta}
\int_0^\Lambda kdk\ {k^2\over 
\kappa(\phi_0) k^2 + V(\phi_0)} = 0.
\label{feme2}
\end{equation}

Now physically we must have $V(\phi_0) > 0$, as for any effective surface tension (it is necessary to have $m$ real), this result implies that  the 
system will naturally be in the region where $\alpha <0$. Now it is 
straightforward to show, see for example~\cite{safran}, at the
the mean field level used here that ratio of the excess area to the projected
area is given as
\begin{equation}
{\Delta A \over A} =   {1\over 4\pi \beta}\int_0^\Lambda kdk\ {1\over 
\kappa(\phi_0) k^2 + V(\phi_0)} = {1\over 8\pi\beta\kappa(\phi_0)}
\ln\left({\kappa(\phi_0)\Lambda^2\over V(\phi_0)} +1 \right).
\end{equation}
In terms of the ratio of the  excess to projected area, equation~\ref{feme2} can now be written as
\begin{equation}
V'(\phi_0)\left[ 1 + {\Delta A\over A}\right] + 
{\kappa'(\phi_0) \over \kappa(\phi_0)}\left[ {\Lambda^2\over 8\pi \beta}
-{V(\phi_0) \Delta A\over A}\right] =0 
\end{equation}
In the limit where $\Delta A/A$ is small using Eq. (\ref{defa}) we obtain
\begin{equation}
\alpha = -{\Lambda^2\over 8\pi \beta V(\phi_0)},
\end{equation}
Now if we write $\Lambda = 2\pi/a$ where $a$ is the microscopic length
scale we find that
\begin{equation}
\alpha = -{\pi k_B T\over 2\epsilon_a}
\end{equation}
where $\epsilon_a$ is the surface energy of a square of the membrane 
of linear dimension $a$, i.e. the average surface energy per lipid. Hence we find $\alpha<0$ which suggests a scenario to observe the formation of mesophases experimentally. For instance one can use a membrane composed by a mixture of DPMC/cholesterol and supported by a frame close at a temperature close to $T_c$. 

In a more general context,
molecular dynamics~\cite{imparato} and Monte Carlo
simulations~\cite{brown} have shown that the bending rigidity has a
non-monotonic behavior as a function of  the short 
lipid number fraction $x_s$: it
first decreases rapidly for small $x_s$ and then increases slowly,
with a minimum around $x_s\simeq 0.6$. These studies suggest that for
a two-component bilayer made of short and long lipids, the gradient of
$\kappa(\phi)$ and $V(\phi)$ could have opposite signs
but some tuning may be required. In this case the effective interaction
will have a repulsive component which could induce mesoscopic phase 
separation.

The issue of mesophase formation has been discussed in several
papers. Taniguchi~\cite{tani} has shown  in a model 
with  a  linear coupling of the composition $\phi$ to the mean 
curvature  that near to spherical vesicles with off-critical 
compositions exhibit circular domains that closely resemble patterns 
observed in red blood cell echinocytosis~\cite{keller}. 

A similar study has been carried  out in 
different geometries~\cite{saxena} and the same general 
phenomena are observed. Inspired by  the problem of  pattern formation 
of quantum dots at the air-water interface, Sear
\textit{et al.}~\cite{sear} have studied the effects of  a 
short-range attraction (on top of a shorter range hard core)
and long-range repulsion in Monte Carlo simulations
of two dimensional systems of interacting particles. In their
simulations both circular domains and stripes were observed as
is the case in the experiments.

Finally, by adding an attractive short-range interaction
to the potential Eq.~(\ref{Phi}), Evans \textit{et al.} have argued that
mesophase formation~\cite{sens} could be induced. Hence, it could explain the
formation of caveolae buds from cell membranes and their striped
texture. The mechanism proposed in this paper of course leads to the 
same phenomenology in the case where the effective potential 
induced by membrane fluctuations has an intermediate range
repulsive component. However  we do not find any repulsion in 
the situation where $\alpha>0$ which implies some conditions on the 
membrane composition which could perhaps be tested experimentally.

The model presented in this paper can be generalized by considering lipid
distributions without the up-down symmetry, i.e. with different
compositions in the up and bottom leaves. In this case, one would
introduce a composition dependent spontaneous curvature, $c(\phi)$ in
the Hamiltonian. If one assumes that the mixed homogeneous phase 
has no spontaneous curvature then one takes  $c(\phi_0)=0$ and in this
case the correction to the long-range interaction is
\begin{equation}
v^*(r)=-\frac{V(\phi_0)}{2\pi}[c'(\phi_0)]^2 K_0(mr)
\end{equation}
and the mass is renormalized (by a repulsive term). Hence this
correction is attractive and could wipe out  the above repulsive
effect. The two-component membrane could also contain
trans-membrane proteins. Despite the fact that the  repulsive interaction
between inclusions described by Evans \textit{et al.} would appear, it
is well known that protein aggregation also increases the local
lipid composition, as observed in erythrocyte membranes where it
induces a phospholipid enrichment~\cite{rodgers}. The inclusion of
protein-like insertions in this two-lipid model could thus produce
quite rich behavior and is a line worth pursuing.

Our study has predicted that it is possible that a membrane whose 
fluctuations are impeded exhibits a macro-phase separation whereas
if it is allowed to fluctuate freely this transition becomes a mesophase
separation . In a stack of membranes the fluctuations are
suppressed by Helfrich forces \cite{helf} which are of  steric origin. 
Experimentally, therefore, one could prepare a stack of bilayers at a lipid
composition where the bilayers within the stack exhibit a
macro-phase separation. However, according to our predictions, a single 
membrane could possibly exhibit a mesophase separation \cite{lau}. 
Another possibility is that one could try and observe the effect predicted 
here by using charged membranes and then varying their rigidity by changing
the bulk solution's salt content \cite{winh}. 

We emphasize that, in this paper, we have concentrated on 
an entirely equilibrium mechanism as a possible explanation for 
the formation of mesoscopic domains. However in living cells,
out of equilibrium effects are of course important. Recently 
the recycling of lipids between the membrane and cell interior has been
put forward  as a non-equilibrium mechanism for the formation of raft-like
structures in active systems~\cite{foret,tuse}.

\end{document}